\documentclass[12pt]{article}
\usepackage{a4}
\usepackage{psfig}
\usepackage{epsfig}
\usepackage{epsf}
\usepackage{rotating}
\usepackage{citesort}       
\usepackage{amssymb}
\usepackage{graphicx}

\newcommand{\vecc}[1]{\mbox{\boldmath $#1$}}
\newcommand{\vR}{\vecc{R}}

\newcommand{\text}[1]{\mbox{$\rm #1 $}}

\newcommand{\anureaction}{\mbox{$\overline{\nu}_e+p\to n+e^{+}$}}

\def\npb#1#2#3{    { Nucl. Phys. }{\bf B #1} (19#2) #3}
\def\npbps#1#2#3{  { Nucl. Phys. }(Proc. Suppl.){\bf B #1} (19#2) #3}
\def\plb#1#2#3{    { Phys. Lett. }{\bf B #1} (19#2) #3}
\def\prd#1#2#3{    { Phys. Rev. }{\bf D #1} (19#2) #3}

\def\prl#1#2#3{    { Phys. Rev. Lett. }{\bf #1} (19#2) #3}

\begin{document}
\begin{flushright}
{\small
IFUM-560/FT\\
FTUAM-02-387
}
\end{flushright}

\vspace{0.2cm}
\begin{center}
{ \Large \bf  KamLAND potentiality on the determination of Neutrino mixing parameters in the post SNO-NC era}\\[0.2cm]
{\large P.~Aliani$^{a\star}$, V.~Antonelli$^{a\star}$, M.~Picariello$^{a\star}$, E.~Torrente-Lujan$^{ab\star}$\\[2mm]
$^a$ {\small\sl Dip. di Fisica, Univ. di Milano},
{\small\sl and INFN Sez. Milano,  Via Celoria 16, Milano, Italy}\\
$^b$ {\small\sl Dept. Fisica Teorica C-XI, 
Univ. Autonoma de Madrid, 28049 Madrid, Spain,}\\
}

\end{center}

\abstract{
We study in detail the power of the 
reactor experiment KamLAND for discriminating existing 
solutions to the SNP and giving accurate information on 
neutrino masses and mixing angles.
Assuming the expected signal 
 corresponding to various ``benchmark'' points in the  
2 dimensional  $(\Delta m^2,\tan^2 \theta)$ mixing plane,
we develop a full-fledged $\chi^2$ analysis which include
KamLAND spectrum and  all the existing solar evidence. 
 A complete modelling of statistical and known
 systematics errors for 1 and 3 years of data taking is 
included, exclusion plots are presented.
  
We find  a much higher sensitivity 
especially for values of $\Delta m^2$ lying in the central 
part of the LMA region. The situation would be more complicate 
for values closer to the border of the LMA region 
(the so called HLMA region, 
i.e. $\Delta m^2 \leq 2 \times 10^{5}$
and $\Delta m^2 \geq 8-9 \times 10^{-5}$ or 
$\tan^2 \theta$ far from $\sim 0.5$). 
In this case kamLAND, with or without solar evidence,
 will only be able  to select 
multiple regions in the parameter space, in the sense that 
different possible 
values of the parameters would produce the same signal.
}

\vskip .5truecm

{PACS: }

\vfill
{\small {}$^\star$ e-mail: paola.aliani@cern.ch, vito.antonelli@mi.infn.it, marco.picariello@mi.infn.it, emilio.torrente-lujan@cern.ch}

\newpage

\section{Introduction}

The publication of the recent SNO 
results~\cite{Ahmad:2002ka,Ahmad:2002jz,howto}
has made an important 
breakthrough towards the solution of the long standing
 solar neutrino 
\cite{Aliani:2002ma,Strumia:2002rv,Bandyopadhyay:2002xj,Barger:2002iv,Pascoli:2002xq,deHolanda:2002pp,Bahcall:2002hv,Foot:2002re,Aliani:2002er}
problem (SNP) possible.
These results provide the strongest evidence so far for flavor oscillation 
in the neutral lepton sector. 

In the near future the reactor experiment KamLAND~\cite{piepke,kamLANDmonaco} is expected 
to further improve our knowledge of neutrino mixing. In fact it should be 
able to sound the region of the mixing parameter space corresponding
to the so called Large Mixing  Angle (LMA) solution of the solar neutrino 
problem  ($\Delta m^2\sim 
10^{-5} - 10^{-4} \ eV^2$ and $\tan^2\theta\sim 10^{-1} - 1 $)
more profoundly.
This is of prime
 interest due to the fact that the LMA region in the one
preferred by the solar neutrino data at present.

The previous generation of reactor experiments 
(CHOOZ~\cite{CHOOZ}, PaloVerde~\cite{PaloVerde}), performed with a baseline of about 1 km.
They have attained  a sensitivity of $\Delta m^2<10^{-3}\ eV^2$ \cite{chooznew,CHOOZ} 
and, not finding any dissapearence of the initial flux, 
they demonstrated that the atmospheric neutrino anomaly
\cite{atmospheric} is not due to muon-electron neutrino 
oscillations. The KamLAND experiment is the 
successor of such experiments at a much larger scale in 
terms of baseline distance and total incident flux.  

This experiment relies upon a 1 kton liquid scintillator
 detector   located at the old, enlarged,  Kamiokande site.
 It searches for the oscillation of antineutrinos 
emitted by several nuclear power plants in Japan. 
The nearby 16  (of a total of 51) nuclear power stations deliver 
a $\overline{\nu}_e$ flux of $1.3\times 10^6 cm^{-2}s^{-1}$
for neutrino energies $E_\nu>1.8$ MeV at the detector position. 
About $78\%$ of this flux comes from 6 reactors forming a 
well defined baseline of 139-214 km. Thus, the flight range 
is limited  in spite of using  several reactors, because of this 
fact the sensitivity of KamLAND will  increase by nearly two 
orders of magnitude compared to previous reactor experiments.

It has been estimated that even in the most conservative 
scenario \cite{piepke}, in 
which the background has to be determined from the reactor 
power fluctuations, the LMA solution is completely within 
the estimated sensitivity after 3 years of data taking. 
Moreover,  KamLAND should be able  \cite{piepke}
to determine  the mixing angle and mass difference with a $20 \%$
accuracy at 99\% of confidence level (CL). 
As has been underlined in~\cite{Petcov:2001sy} and 
later in ~\cite{HLMA},
the problems might arise if the value of the 
squared mass difference ($\Delta m^2$) lies in the upper 
region of the LMA 
solution, the so called HLMA region.

Let us briefly recall some  model independent conclusions
obtained from the  results of  SNO~\cite{Aliani:2002ma}.
 Different quantities can be defined in order to make  
the evidence for 
disappearance and  appearance of the neutrino flavors explicit. 
Letting alone the SNO data, from the three fluxes measured 
by SNO is possible to 
define two useful ratios: 
$\phi_{CC}/\phi_{ES}$, $\phi_{CC}/\phi_{NC}$, 
deviations of these ratios with respect to  their 
standard value are powerful tests for occurrence of new physics.
For the first ratio, one  obtains~\cite{Aliani:2002ma}
$$ \frac{\phi_{CC}}{\phi_{ES}}=0.73^{+0.10}_{-0.07}, $$
 a value which is  $\sim$ 2.7 $\sigma$ away from the 
 no-oscillation expectation value of one.
The ratio of CC and NC fluxes gives the fraction of
 electron neutrinos remaining 
in the solar neutrino beam, the  value obtained in 
Ref.~\cite{Aliani:2002ma}:
$$\frac{\phi_{CC}}{\phi_{NC}}=0.34^{+0.05}_{-0.04},$$
this value is far away  from the 
 standard model case.

Finally, if in addition to SNO data one consider the 
 flux predicted by the solar standard mode one can define, 
following   Ref.\cite{Barger:2001zs},  the 
quantity $\sin^2\alpha$, the fraction of 
'' neutrinos which oscillated into active ones'', 
one finds the following result:
$$
\sin^2\alpha=\frac{\Phi_{NC}-\Phi_{CC}}{\Phi_{SSM}-\Phi_{CC}}=0.92^{+0.39}_{-0.20},
$$
where the SSM flux is taken as the ${}^8\rm B$ flux predicted 
in Ref.\cite{bpb2001}.
The central value is clearly below one (only-active oscillations).
Although electron neutrinos are still allowed to oscillate into 
sterile neutrinos the hypothesis of transitions to {\em only} sterile 
 neutrinos is rejected at nearly $5\sigma$.

The aim of this work is to study the KamLAND 
discriminating power, to understand in which regions
 of the parameter space still allowed by the solar 
neutrino experiments KamLAND might give satisfactory accuracy.
The structure of this work is the following.
In section \ref{kamland} we discuss the main features of KamLAND experiment 
that are relevant for our analysis:
we derive updated numerical expressions for the 
reactor fuel cycle-averaged antineutrino flux and the
absortion antineutrino cross section.
 The next section is devoted to the 
salient aspects of the procedure we are adopting.
The results of our analysis are presented and discussed in
 section 
\ref{analysis} and, finally, in section \ref{conclusions} we draw our 
conclusions and discuss possible future scenarios. 

\section{A Kamland overview}
\label{kamland}

Electron antineutrinos from nuclear reactors  with energies 
above 1.8 MeV are measured in KamLAND by detecting the inverse 
$\beta$-decay reaction $\overline{\nu}_e+p\to n+e^+$. The time 
coincidence, the space correlation and the energy balance  
between the positron signal and the 2.2 MeV $\gamma$-ray
 produced by the capture of a already-thermalized  neutron on a
 free proton make it possible to identify this reaction 
unambiguously, even in the presence of a rather large background. 

The two principal ingredients in the calculation of the expected 
signal in KamLAND are  the reactor  flux and
the antineutrino cross section on protons. These 
ingredients are considered below. 

\subsection{The reactor antineutrino  flux}

We first describe the flux of antineutrinos coming from 
the power reactors.
A number of short baseline experiments 
(~Ref.\cite{Murayama:2000iq} and references therein) 
have measured the energy spectrum of reactors at distances 
where oscillatory effects have been shown to be inexistent. 
They have shown that the theoretical neutrino flux predictions 
are reliable within 2\% \cite{piepke}.

The effective flux of antineutrinos released by the nuclear
 plants is a rather well  understood function of 
the thermal power of the reactor and
 the amount of thermal power emitted during the 
fission of a given nucleus, which gives the total amount, and 
the  isotopic composition of the reactor fuel which gives the 
spectral shape.
Detailed tables for these
 magnitudes can be found in Ref.~\cite{Murayama:2000iq}.

For a given isotope $(j)$ the energy  spectrum can be parametrized 
by the following expression  \cite{vogel}
\begin{eqnarray}
d N_\nu^{j}/d E_\nu=\exp (a_0+a_1 E_\nu+ a_2 E_\nu^2)
\label{reactorspectrum}
\end{eqnarray}
where the coefficients $a_i$ depend 
on the nature of the fissionable isotope 
(see Ref.\cite{Murayama:2000iq} for explicit values).
Along the year, between periods of refueling, the total effective flux changes with time as the fuel is expended and the isotope 
relative composition varies.
The overall spectrum is at a given time
$$ \frac{d N_\nu}{d E_\nu}=\sum_{j=isotopes} 
c_j(t)\frac{d N_\nu^{j}}{d E_\nu}.$$
To compute a fuel-cycle averaged spectrum
we have made use of the typical evolution of the relative 
abundances $c_j$, which  can be seen in Fig. 2 of 
Ref.\cite{Murayama:2000iq}.
This averaged spectrum can  be again  fitted very well by 
the same functional expression (\ref{reactorspectrum}).
The isotopic energy yield  is properly taken into account. 
As the result of this fit, we obtain 
the following values which are the ones to be used in the 
rest of this work:
$$a_0=0.916,\quad a_1=-0.202,\quad a_2=-0.088.$$   

Although individual variations of the $c_j$ along the 
fuel cycle can be very high, the variation of the two most 
important ones is highly correlated: the 
coefficient $c({}^{235} U)$ increases in the range
 $\sim 0.5-0.7$ while 
$c({}^{239} Pu)$ decreases $\sim 0.4-0.2$. 
This correlation makes  the effective description of the 
total spectrum by a single expression Eq.\ref{reactorspectrum} useful.
With the fitted coefficients $a_i$ above, the difference between
this effective spectrum and the real one is typically $2-4\%$ 
along the yearly fuel cycle.

\subsection{Antineutrino cross sections}

We now consider the
cross sections for antineutrinos on protons. We will 
sketch the form of the well known differential expression and 
more importantly we will give updated numerical values 
for the transition matrix elements which appear as 
coefficients. 

In the limit of infinite nucleon mass, the 
cross section  for the reaction 
\anureaction\  is given by \cite{zacek,reines} 
$\sigma(E_{\overline{\nu}})=k E_{e^+} p_{e^+}$
where $E,p$ are the  positron energy and momentum
 and $k$ a  transition matrix element which will be 
considered 
below.
The  positron spectrum  is  monoenergetic: 
 $E_{\overline{\nu}}$ and  
$E_{e^+}$ are related by:
$E_{\overline{\nu}}^{(0)}=E_{e^+}^{(0)}+\Delta M$,
where $M_n, M_p$ are the neutron and proton masses
 and $\Delta M=M_n-M_p\simeq 1.293 $ MeV.

Nucleon recoil corrections are 
potentially important in relating the positron and antineutrino 
energies in order to evaluate the antineutrino flux. 
Because the antineutrino 
flux $\Phi(E_\nu)$ would typically decrease quite 
rapidly with energy, the lack of adequate corrections 
will systematically overestimate the   positron yield.

At highest orders,  the  positron spectrum  is not 
monoenergetic and one has to integrate over the positron angular 
distribution to obtain the positron yield.
The  differential cross section  at first order $1/M_p$ is of 
the form:
\begin{eqnarray}
\left(\frac{{\rm d}\sigma}{{\rm d cos}\theta}\right)^{(1)}
& = & \frac{\sigma_0}{2}
\left[\vphantom{\frac{\sigma_0}{2}}
(f^2 + 3g^2) + (f^2 - g^2) v_e^{(1)} \cos\theta\right]
E_e^{(1)} p_e^{(1)} \nonumber 
\label{eq:dsig1}
\end{eqnarray}
Complete expressions and notation can be found 
in Ref. \cite{vogel}. Here we only want to pay attention to 
the overall coefficient $\sigma_0$ which is related to the 
transition matrix element $k$ above.

The matrix transition element can be written in terms of 
measurable quantities as 
$$k=2\pi^2 \log 2/ (m_e^5 f \, t_{1/2})$$
where appears the free neutron
decay $t_{1/2}$, the phase-space factor f and $m_e$.
The value of  $f=1.71465\pm 0.00015$ follows from calculation 
\cite{zacek21},
 while $t_{1/2}=613.9\pm 0.55$ 
sec is the latest published value for the neutron half-life \cite{PDG2002}. This value has a  significantly 
smaller  error than previously quoted measurements.

From the values above, we obtain the extremely precise value: 
$$k=(9.5305\pm 0.0085)\times 10^{-44}\  cm^2/MeV^{2}.$$ 
From here the coefficient which appears in the differential 
cross section is obtained as
(vector and axial vector couplings $f=1,g=1.26$):
$ k=\sigma_0 (f^2+3 g^2).$
In summary, the differential cross section which appear in
  KamLAND are very well known, its theoretical 
errors are negligible if updated values are employed.

\section{The computation of the expected signals}
\label{klsignal}

In order to obtain the
 expected number of events at KamLAND, we sum the 
expectations for all the relevant reactor sources weighting 
each source by its power and distance to the detector
(table II in Ref.~\cite{Murayama:2000iq} ), 
 assuming the same 
spectrum originated from each reactor. The
average number of positrons $N_i$ which are detected per
visible energy bin $\Delta E_i$ is given by the convolution of 
different quantities: 
$\overline{P}$, the oscillation probability 
averaged  over the  distance and power of the different 
reactors, the antineutrino capture cross section
  given as before,
the antineutrino flux spectrum  given by
 expression \ref{reactorspectrum}, the relative reactor-reactor 
power normalization which 
is included in the definition of 
 $\overline{P}$ and 
the energy resolution of KamLAND which 
is rather good \cite{kamLANDmonaco}, 
we use in our analysis the expression  
$\sigma(E)/E\sim 10\%/\surd E$.

For one year of running with the 600 ton fiducial 
mass and for standard 
nuclear plant power and fuel schedule: we assume all the 
reactors operated at $\sim 80\%$ of their maximum capacity
and an averaged, time-independent, fuel composition 
equal for each detector, the experiment expects about 550 
antineutrino events (this number agrees with 
 other estimations. i.e. Ref.\cite{kamLANDmonaco}). 
We will consider 
this number as our KamLAND year in what follows.
We will not add any background events as we will 
suppose that they can be distinguished from the signal 
with sufficiently high efficiency. This should be taken 
with caution,  
we will dedicate a few words to the experimental 
background at the end of section.

We compute the expected signal in 
 a set of 0.5 MeV width 
total positron energy bins in the 
range 2.0 MeV - 8 MeV. 
Quantitavely,
the main information content of the shape of the observable 
spectrum is
summarized by the first moment of the distribution, the average 
spectrum energy. 
This first moment is defined as
\begin{equation}
\langle E\left(\theta, \Delta m^{2}\right)\rangle=
\frac{\sum_i \overline{E}_{i}R_{i}\left(\theta, 
\Delta m^{2}\right)}{\sum_i R_{i}\left(\theta, \Delta m^{2}\right)}
\end{equation}
where $\overline{E}_{i}$ is the center positron energy in 
the i$^{\text{th}}$ energy bin 
and  the normalized signals relative to the
non-oscillation case: 
$R_i=N_i(\theta,\Delta m^2)/N_i^0$.

\begin{figure}
\centering
\begin{tabular}{lr}
\psfig{file=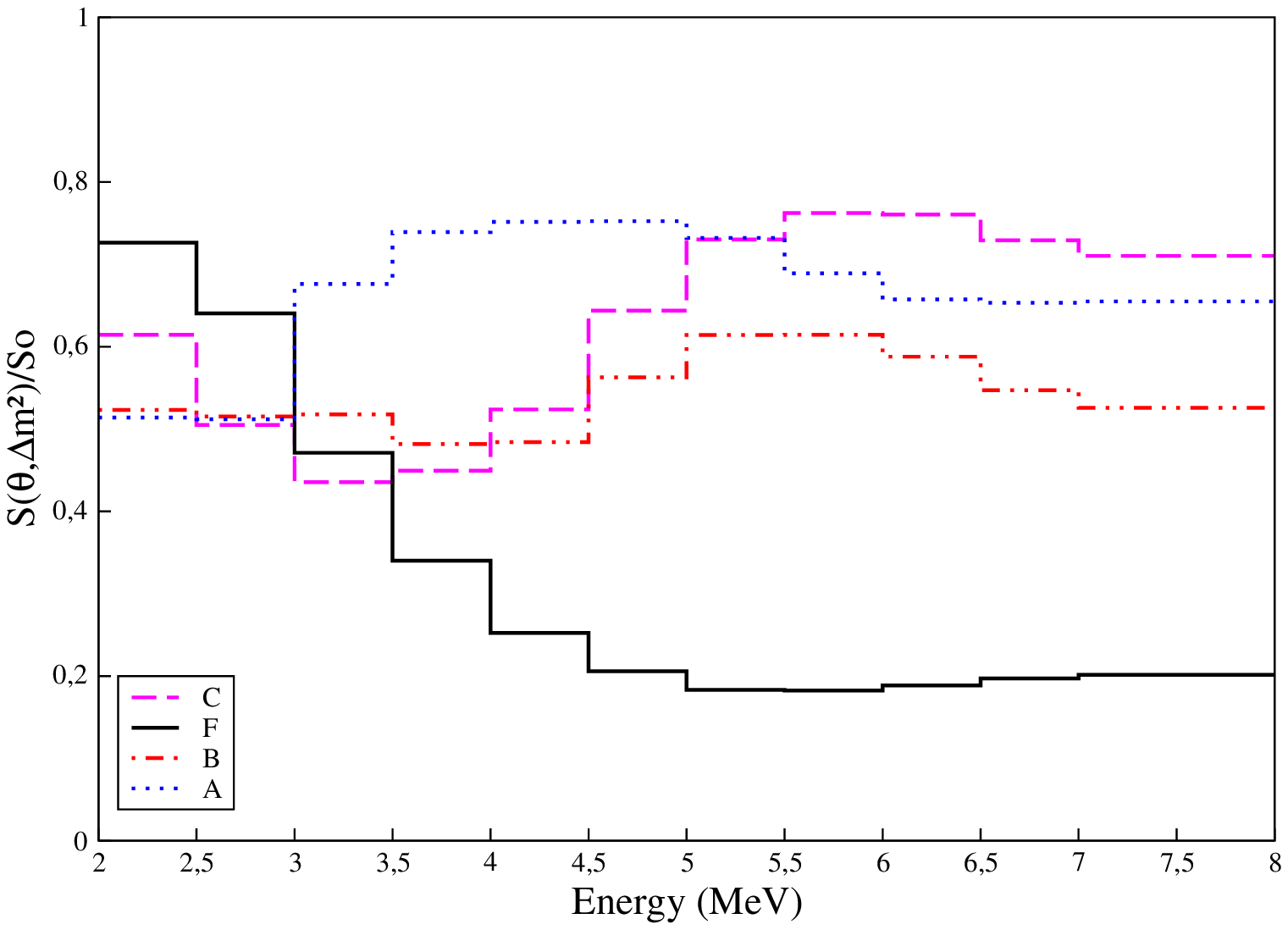,width=7cm}&
\psfig{file=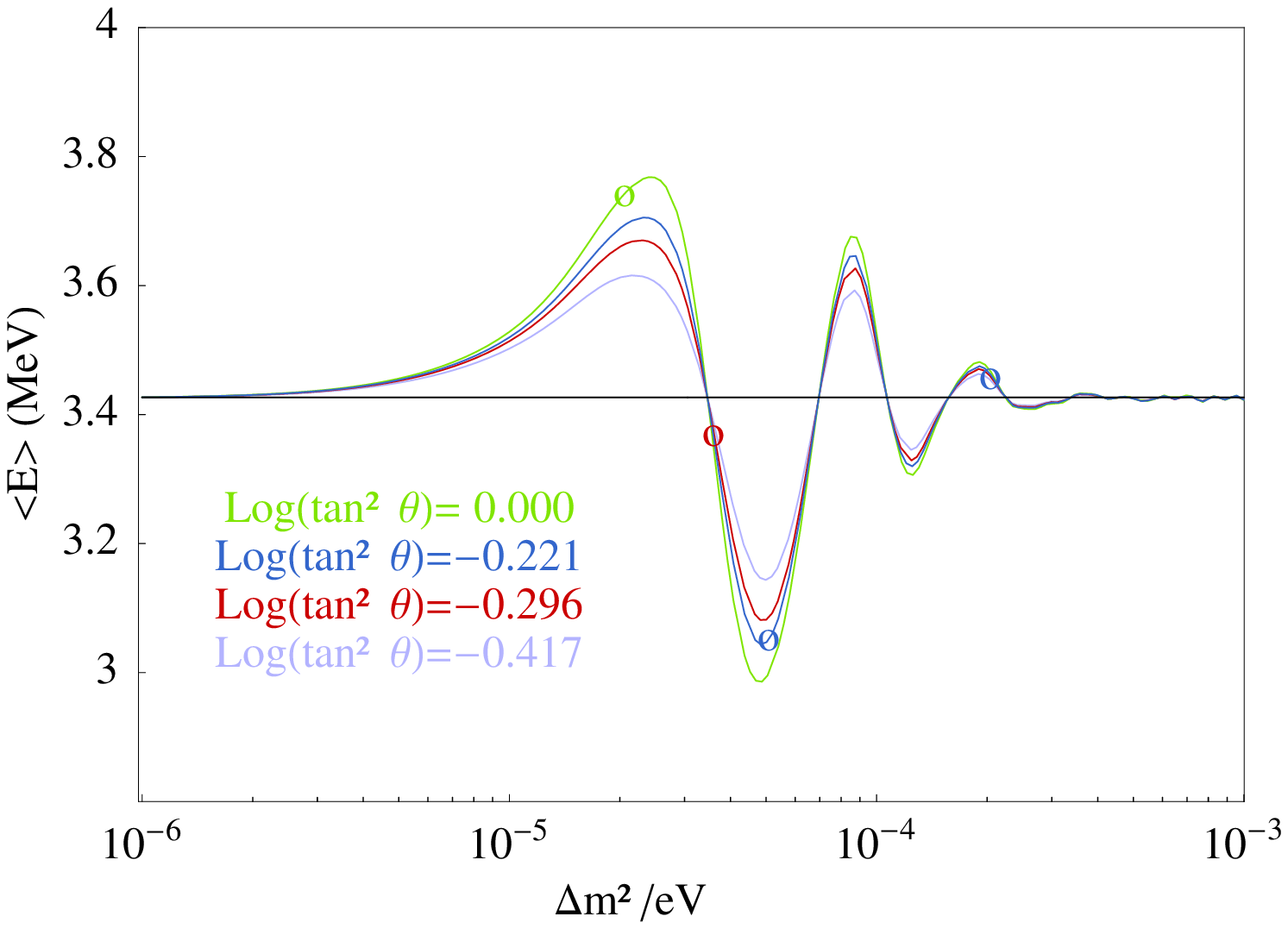,width=9.5cm}
\end{tabular}
\caption{(Left) Different kamLAND spectra for points in
 the LMA region.
(Right) KamLAND moments distribution as a function of the
 oscillation parameters. 
The No-oscillation mean-spectrum 
energy is given by the horizontal black line.}
\label{moments}
\end{figure}

The expected value of this first moment as a function of 
$\Delta m^2$ for some selected values of the mixing angle 
is represented in Fig. \ref{moments}(right). 
We want to illustrate the potentiality of this 
magnitude as an indicator of neutrino oscillations at 
KamLAND. In this plot we graphically see the fact that KamLAND 
is  sensitive to neutrino oscillations in the LMA region and in 
particular to the $\Delta m^{2}$ parameter. 
The first moment can vary up $\sim 12\%$ in
sizeable regions of the parameter space (this can be beautifully
 seen  in a 3D plot). A variation of a fraction of this size 
should be clearly identifiable from the KamLAND data after 
 1 year of effective running.
In order to see whether this result depends on the choice 
of bin-size, we have also reproduced it with larger, 1 MeV bins, 
and found that there is no significant difference. 

In Fig.\ref{moments} (left) we show the visible positron 
energy spectrum at KamLAND for some chosen oscillation 
parameters (see Table \ref{t1}). 
We present the integrated signal at every 0.5 MeV bin
 normalized to the non-oscillation expectation.
We can see from the plot how the shape of the signal is 
very sensitive to the oscillation parameters. It can greatly
 change through the LMA region. From the point of view of
background reduction, in some favorable cases the 
spectrum is peaked above 5 MeV, this suggests the extension 
of the fiducial energy thresholds beyond the 8 MeV level.

In addition to the  antineutrino 
 signal, two classes of background can 
be distinguished 
\cite{piepke,Murayama:2000iq,brackeeler}.

The random coincidence background is due to the contamination of 
the detector scintillator by U, Th and Rn. 
From MC studies and assuming that an adequate level of
purification can be obtained, 
 the background coming from this source
is expected to be $\sim 0.15$ events/d/kt which is 
equivalent to a signal to 
background ratio of  $\sim 1\%$. 
Other works \cite{usareport} conservatively estimate 
a $5\%$ level for this ratio.   
More importantly for what it follows, one expects that the 
random coincidence  backgrounds will be a relatively steeply 
falling function of energy. The assumption of no background 
should be relatively safe only 
at high energies (above $\sim 5$ MeV).

The second source of background,
the so called correlated background is dominantly caused by 
cosmic ray muons and neutrons. The KamLAND's depth is the main 
tool to suppress those backgrounds. MC methods estimate a 
correlated background of around $0.05$ events/day/kt 
distributed 
over all the energy range up to $\sim 20$ MeV.

We will also  need  the expected signals in the 
different solar neutrino experiments.
These  are obtained by convoluting solar neutrino 
fluxes, sun  and earth 
oscillation probabilities, neutrino cross sections 
and detector energy response functions. We closely follow 
the same methods already well explained in previous works
\cite{Aliani:2001zi,Aliani:2001ba,Aliani:2002ma,Aliani:2002rv}, 
we will mention here only a few aspects 
of this computation.
We determine the neutrino oscillation probabilities 
using the standard
methods found in literature~\cite{torrente}, as explained 
in detail in~\cite{Aliani:2001zi} and in~\cite{Aliani:2002ma}. 
We use a thoroughly numerical method to calculate the 
neutrino evolution equations in the presence of matter for all 
the parameter space.
For the solar neutrino case 
the calculation is split in three steps, corresponding to 
the neutrino propagation inside the Sun, in the vacuum 
(where the propagation is computed analytically) and in the Earth.
We average over the neutrino production point inside the Sun 
and we take the electron number density $n_e$ in the Sun by the BPB2001 model~\cite{bpb2001}.
The averaging over the annual  variation of the orbit
is also exactly performed.
To take the Earth matter effects into account, we adopt a
 spherical model of the Earth  density and chemical composition~\cite{earthprofile}.
The gluing of the neutrino propagation in the three different 
regions is performed exactly using an evolution operator
formalism~\cite{torrente}.
The final survival probabilities are obtained from the 
corresponding (non-pure) density matrices built from 
the evolution operators in  each of these three regions.

In this analysis in addition to night probabilities we will 
need the  partial night probabilities corresponding 
to the 6 zenith angle bin data presented by SK \cite{Smy:2002fs}.
They are obtained using appropriate weights 
 which depend on the neutrino  impact parameter and the 
 sagitta distance from neutrino trajectory  to the Earth's center,
 for each detector's geographical location.

\section{Analysis and Results}
\label{analysis}

In order to study the potentiality of KamLAND for 
resolving the neutrino oscillation parameter space,
we have developed  two kind of analysis. In the first 
case (Analysis A below) we will  deal with the KamLAND 
expected global signal. We will asumme that the experiment 
measure a certain global signal 
with given statistical and systematic 
error after some period of data taking 
(1 or 3 yrs) and will perform a complete $\chi^2$ 
statistical analysis including in addition the 
up-date solar evidence. 
In the second case, Analysis B, we will include the full 
KamLAND spectrum information. 
Instead of giving arbitrary values to the different bins, 
we will assume a number of oscillation models characterized 
by their mixing parameters $(\Delta m^2,\theta)$. 
After including the  solar evidence 
we  will perform the same $\chi^2$ analysis as before.  

\subsection{Analysis A}

The total $\chi^2$ value is given by the sum of 
two distinct contributions, that is the one coming from all 
the solar neutrino data and the contribution of the KamLAND 
experiment:
$$  \chi^2= \chi^2_{\odot} + \chi^2_{glob,KL},$$
with
\begin{eqnarray}
  \chi^2_{glob,KL}&=& \left 
(\frac{R^{exp}-R^{teo}(\Delta m^2,\theta)}{\sigma_{stat+sys}}\right )^2.
\end{eqnarray}
For the ``experimental'' signal ratio $R^{exp}$ we assume 
different values varying from a very strong suppression 
$R\sim 0.3$ to unobservation of neutrino oscillations 
$R\sim 1$.
The total error $\sigma$ is computed as a sum of assumed 
systematics deviations, $\sigma_{sys}/S\sim 5\%$,
 mainly coming from flux uncertainty 
($3\%$), energy baseline calibration and others 
(see Ref.\cite{klsystematics,Murayama:2000qi,kamLANDmonaco}) 
and statistical errors $\sigma_{stat}\sim \surd S$. 
The   nominal periods of data taking that we 
consider, 1 and 3 yrs, are generical representative  cases 
where systematical or
statistical errors are taken as predominant.


The solar neutrino contribution can be written in 
the following way:
\begin{eqnarray}
  \chi^2_{\odot }&=& \chi^2_{\rm glob}+\chi^2_{\rm SK}+\chi^2_{\rm SNO}.
\end{eqnarray}
The function $\chi^2_{\rm glob}$ 
correspond to the total event rates measured at the  
Homestake  experiment~\cite{Homestake} and at the gallium 
experiments SAGE~\cite{sage,sage1999}, 
GNO~\cite{gno2000} and GALLEX~\cite{gallex}. 
We follow closely the definition used in 
previous works (see Ref.\cite{Aliani:2002ma}
 for definitions and 
 Table~(1) in Ref.\cite{Aliani:2002ma} ` 
for an explicit list of results 
and other  references).

The contribution to the $\chi^2$ from the SuperKamiokande 
data ($\chi^2_{\rm SK}$) has been obtained by using  
  double-binned data in energy 
 and zenith angle (see table 2 in Ref.\cite{Smy:2002fs} and
 also Ref.\cite{Fukuda:2002pe}):
8 energy bins of variable width and 7 zenith angle bins which
 include the day bin and 6 night ones. The definition is 
given by: 
\begin{eqnarray}
\label{chi2}
  \chi^2_{\rm SK}
&=& ({\alpha \vR^{\rm th} -\vR^{\rm exp}})^t 
\left (\sigma^{2}_{\rm unc} + \sigma^{2}_{\rm cor}\right )^{-1}
 ({\alpha \vR^{\rm th}-\vR^{\rm exp}}).
\end{eqnarray}
The theoretical and experimental $\vR$ quantities  
are this time matrices of dimension 8$\times$7. 
The factor $\alpha$ is a flux normalization with respect to 
the value measured by SNO NC. 
The covariance quantity $\sigma$ is a 4-rank 
tensor constructed  in terms of
statistic errors, energy and zenith angle 
bin-correlated and uncorrelated uncertainties.
The data and errors for individual energy 
 bins for SK spectrum has been obtained from Ref.~\cite{Smy:2002fs}. 

The contribution of SNO to the $\chi^2$ is given by
\begin{eqnarray}
  \chi^2_{\rm SNO }&=&
\left( \frac{\alpha - \alpha^{\rm th}}{\sigma_\alpha}\right)^2 +
 \chi^2_{\rm spec-SNO }
  \label{chiall}
\end{eqnarray}
The presence of the first term is due to the introduction  in 
our analysis
of the flux normalization factor $\alpha$ with respect to the 
SNO NC flux, 
whose central and error values are given in 
Table I of Ref.\cite{Aliani:2002ma}. 
The second term in formula \ref{chiall} corresponds to:
\begin{eqnarray}\label{chi2b}
  \chi^2_{\rm spec-SNO}
&=&\sum_{d,n} ({\alpha \vR^{\rm th} -\vR^{\rm exp}})^t 
\left (\sigma^{2}_{\rm stat} + \sigma^{2}_{\rm syst}\right )^{-1}
 ({\alpha \vR^{\rm th}-\vR^{\rm exp}}),
\end{eqnarray}
where the day and night $\vR$ vectors of dimension 17 
are made up by the
values of the total (NC+CC+ES) SNO signal for the 
different bins of the spectrum.  
The statistical contribution to the covariance 
matrix, $\sigma_{\rm stat}$,
is obtained directly from the SNO data. The part of the matrix 
related to the systematical errors  has been computed  
by us studying the influence on the response function 
of the different sources of 
correlated and uncorrelated errors reported by SNO 
collaboration (see table II of
 Ref.~\cite{Ahmad:2002jz}), we assume 
full correlation or full anticorrelation according to 
each source.

To test a particular oscillation hypothesis against the 
parameters of the best fit 
and obtain allowed regions in parameter space we perform a 
 minimization of the three dimensional function
 $\chi^2(\Delta m^2,\tan^2\theta,\alpha)$. 
For $\alpha=\alpha_{\rm min}$, 
 a given point in the oscillation parameter space is allowed if 
 the globally subtracted quantity 
fulfills the condition 
 $\Delta \chi^2=\chi^2 (\Delta m^2, \theta)-\chi_{\rm min}^2<\chi^2_n(CL)$.
Where $\chi^2_{n=3}(90\%,95\%,...)$ are the quantiles for
 three degrees of freedom.

\begin{figure}
\centering
\begin{tabular}{rl}
\psfig{file=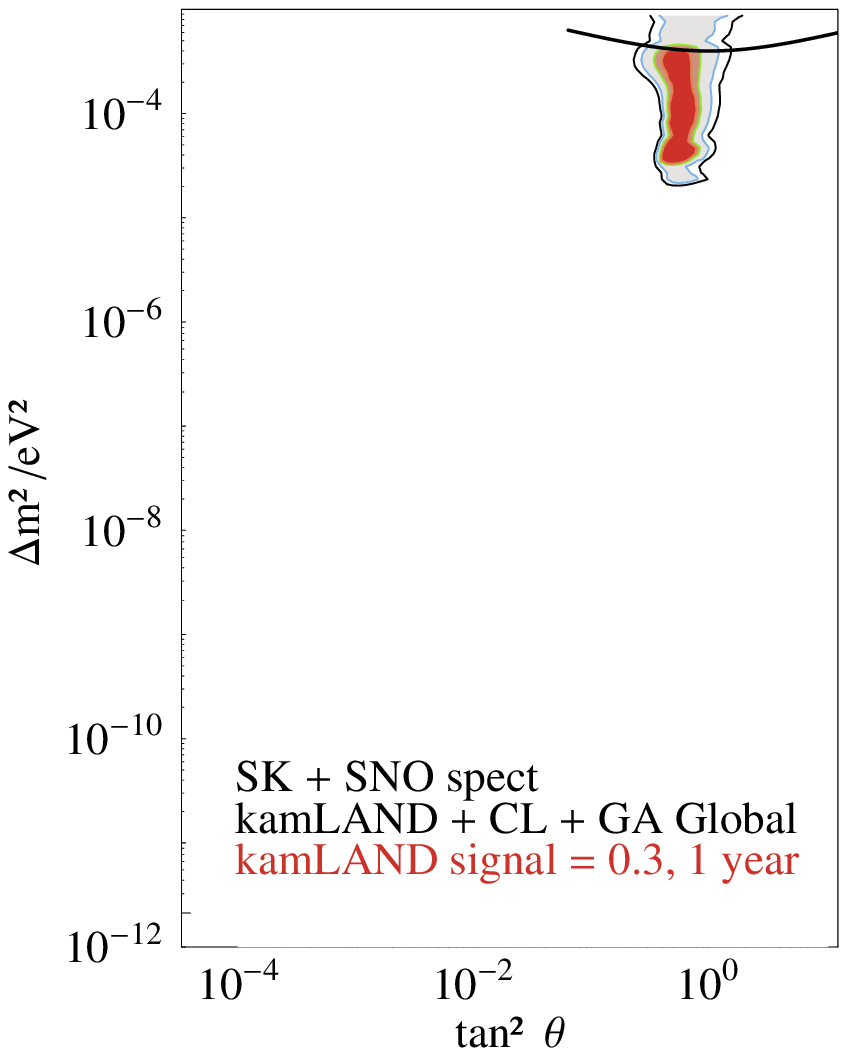,width=5cm}&\psfig{file=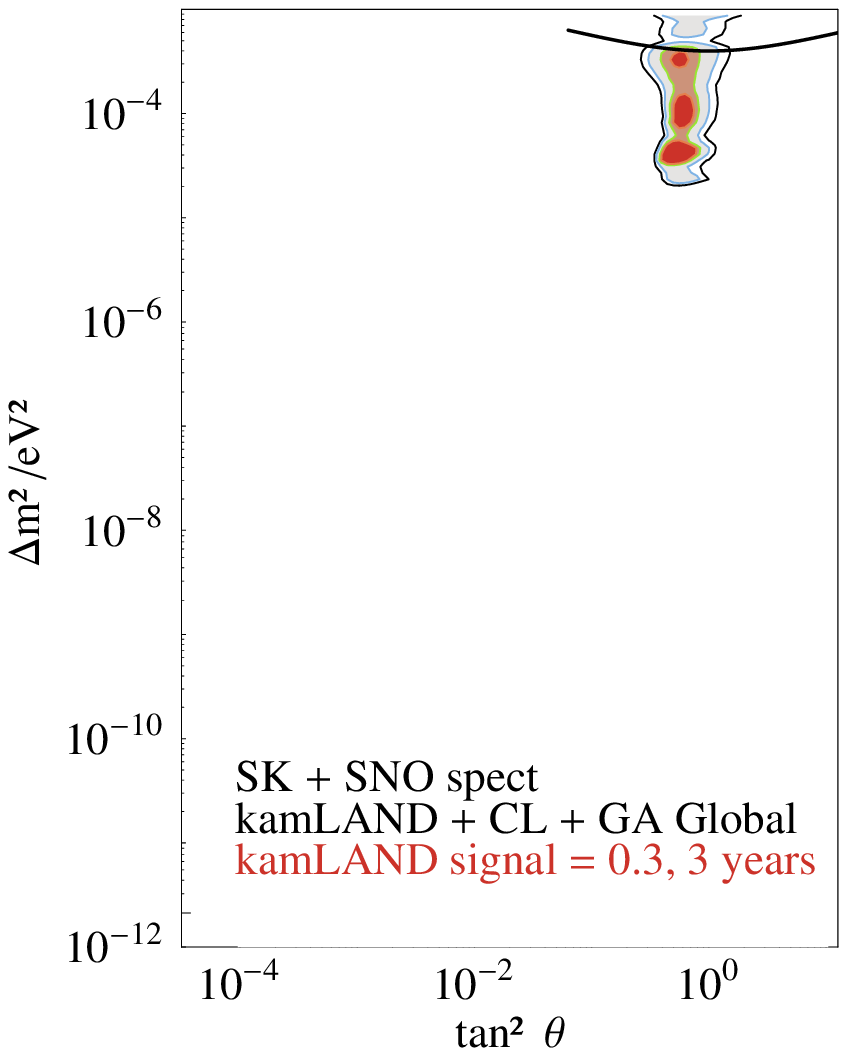,width=5cm}\\[0cm]
\psfig{file=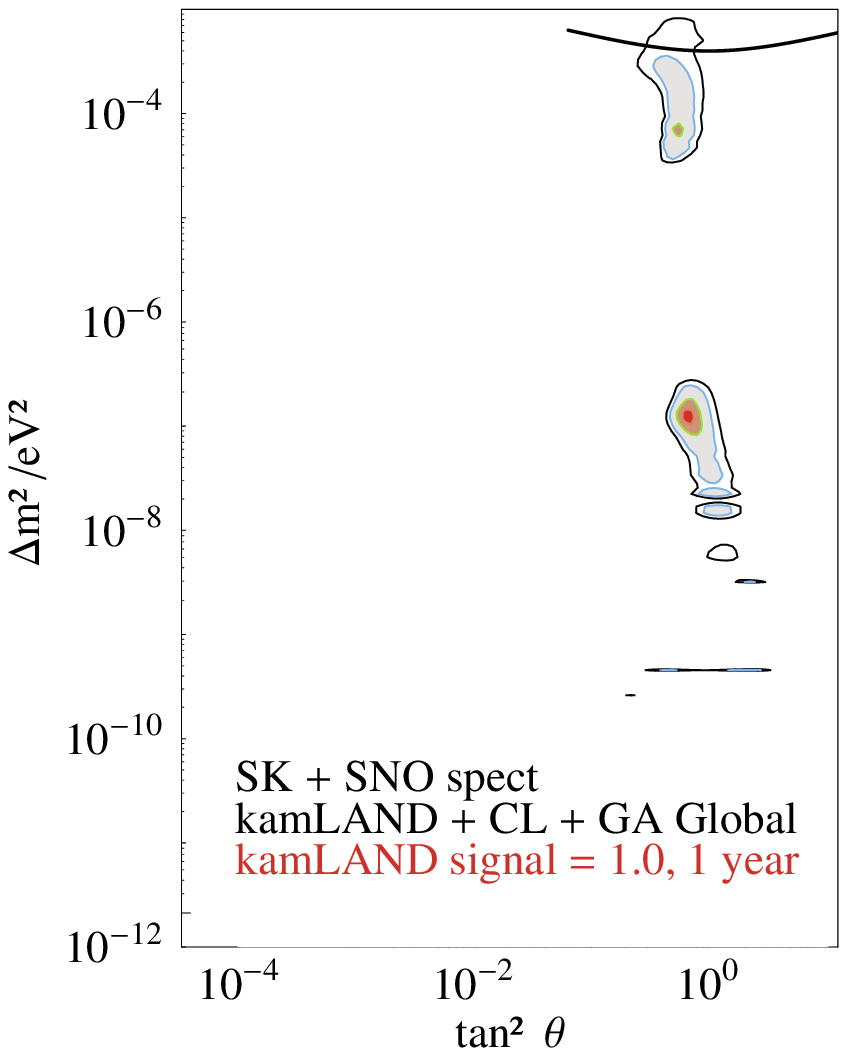,width=5cm}&\psfig{file=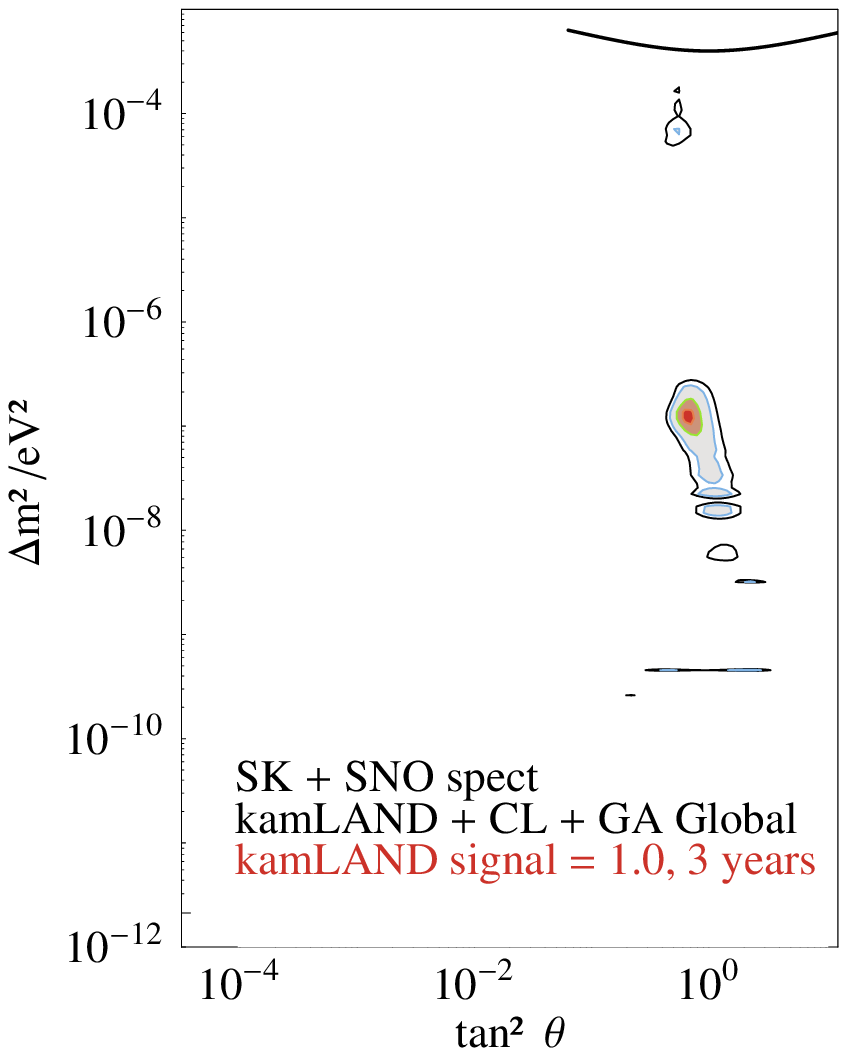,width=5cm}\\[0cm]
\end{tabular}
\caption{
\small Exclusion plots including KamLAND global rates (Analysis A), given a
 hypothetical experimental global signal ratio: 
 respectively  $S/S_0= 0.3,0.6$, and no oscillation 
 evidence $S/S_0= 1.0$. 
Statistical and assumed systematics 
($\sim 5\%$) errors are included.  
left (right) panel refers to one (three) year of KamLAND 
data taking.
The colored areas are allowed   at 
90, 95, 99 and 99.7\% CL relative to the absolute minimum.
The region above the upper thick line is excluded by the 
reactor experiments \protect\cite{chooznew}.
}
\label{f1}
\end{figure}

In Fig.\ref{f1} we graphically show the 
results of this analysis. They represent
exclusion plots including KamLAND global rates, given a
 hypothetical experimental global signal ratio: 
 respectively strong and medium suppression $S/S_0= 0.3,0.6$ 
and no oscillation 
 evidence $S/S_0= 1.0$ for one and three years of KamLAND 
data taking.
As can be seen in the figures, as the 
KamLAND experimental signal decreases, the LMA region is singled 
out. The periodic shape in $\Delta m^{2}$ of the 90 \% C.L. 
(red regions) which becomes apparent in the three-year plot 
 is due to the periodicity of the 
response function: in order to distinguish among these different 
equally-likely solutions, one
 must analyze the energy spectrum, this
will done in the main analysis to be presented in the next 
sections. 
Obviously, only if KamLAND sees some oscillation signal 
(i.e. $S_{i}/S_{0} << 1.0$ ) does the LMA region 
become the only solution. If we consider a hypothetical signal 
closer to 1.0 than 0.3, we see that the LOW region survives, 
although it is less favored.

\subsection{Analysis B}

Here we use the expected binned KamLAND signal for 
some benchmark, arbitrarily chosen,  points
in parameter space that we show in table (\ref{t1}).
For any of these points we obtain the expected spectrum 
after 1 or 3 years of data taking under the ``standard'' 
conditions described above. Next we perform an standard 
 $\chi^2$ analysis introducing statistical and assumed 
 systematics errors including the evidence of the up-date 
 solar experiments (CL,GA,SK and SNO).

In the present study the total $\chi^2$ value is given by the 
sum of two distinct contributions, that is the one coming from 
all the solar neutrino data and the contribution of the 
KamLAND experiment:
\begin{eqnarray}
  \chi^2&=& \chi^2_{\odot} + \chi^2_{spec,KL}.
\end{eqnarray}
The contribution of the solar neutrino experiments
$\chi^2_{\odot}$
 is described in detail in the previous section.
The contribution of the KamLAND experiment is now as follows:
\begin{eqnarray}
  \chi^2_{\rm spec,KL}&=& ({\vR^{\rm th,0}-\vR^{\rm th}})^T 
\left (\sigma_{unc}^2+\sigma_{corr}^2 \right)^{-1} ({\vR^{\rm th,0}-\vR^{\rm th}})
\end{eqnarray}
Note that the addition to $\chi^2$ of a constant 
term $N_{dof}$ has
no practical importance for the main purpose of this kind
of  analysis: the determination of exclusion regions proceeds 
from the  minimum subtracted quantity $\chi^2-\chi^2_{min}$.
The ${\vR}$ are length 12 vectors containing the binned 
spectrum (0.5 MeV bins ranging from 2 to 8 MeV) 
normalized to the non-oscillation 
 expectations. Theoretical vectors are a function of the 
oscillation parameters:
$\vR^{\rm th}=\vR^{\rm th}(\Delta m^2,\theta)$. 
 The ``experimental'' vectors
 are defined in similar way, for any of the benchmark points 
$(\Delta m^2_0,\theta_0)$ we have 
$\vR^{\rm th,0}=\vR^{\rm th,0}(\Delta m^2_0,\theta_0)$.

We generate acceptance contours
 (at 90,95 and 99 \% CL)  in the 
$(\Delta m^2,\tan^ \theta)$ plane in a similar manner 
as explained in the previous section. For the sake of 
comparison we have also obtained exclusion regions derived 
from the consideration of the $\chi^2_{spec,KL}$ alone.

\begin{figure}
\centering
\begin{tabular}{lr}
\psfig{file=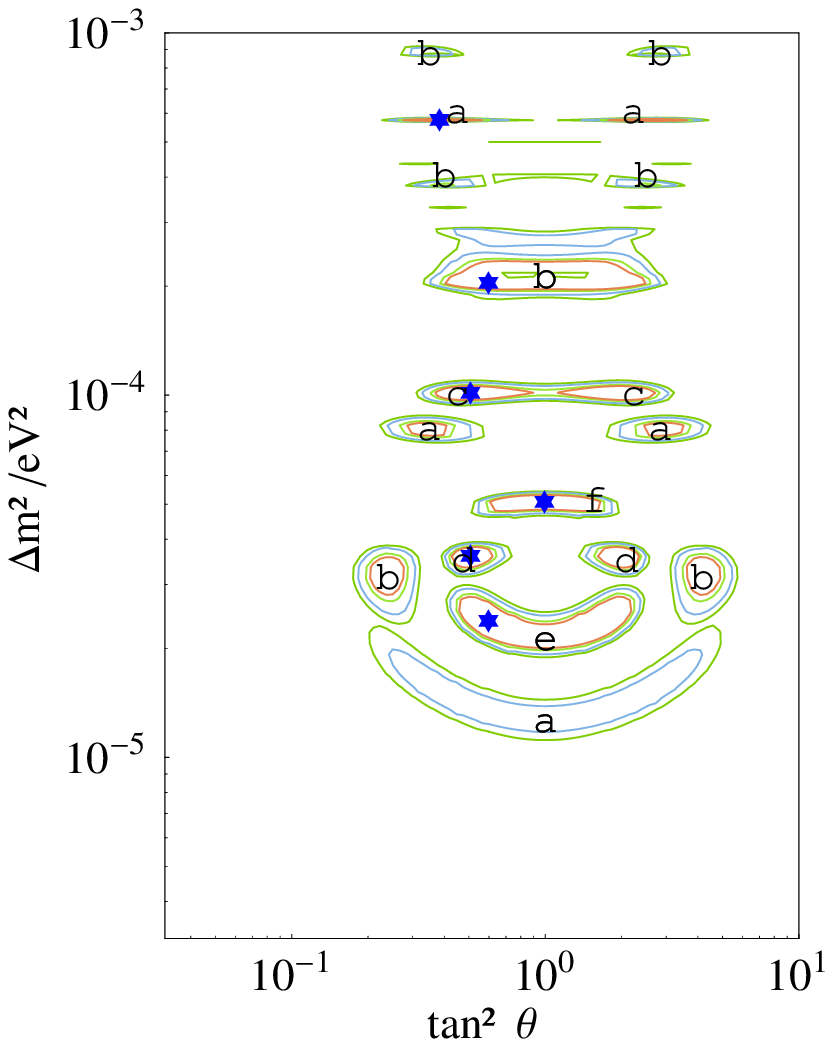,width=8cm}
&\psfig{file=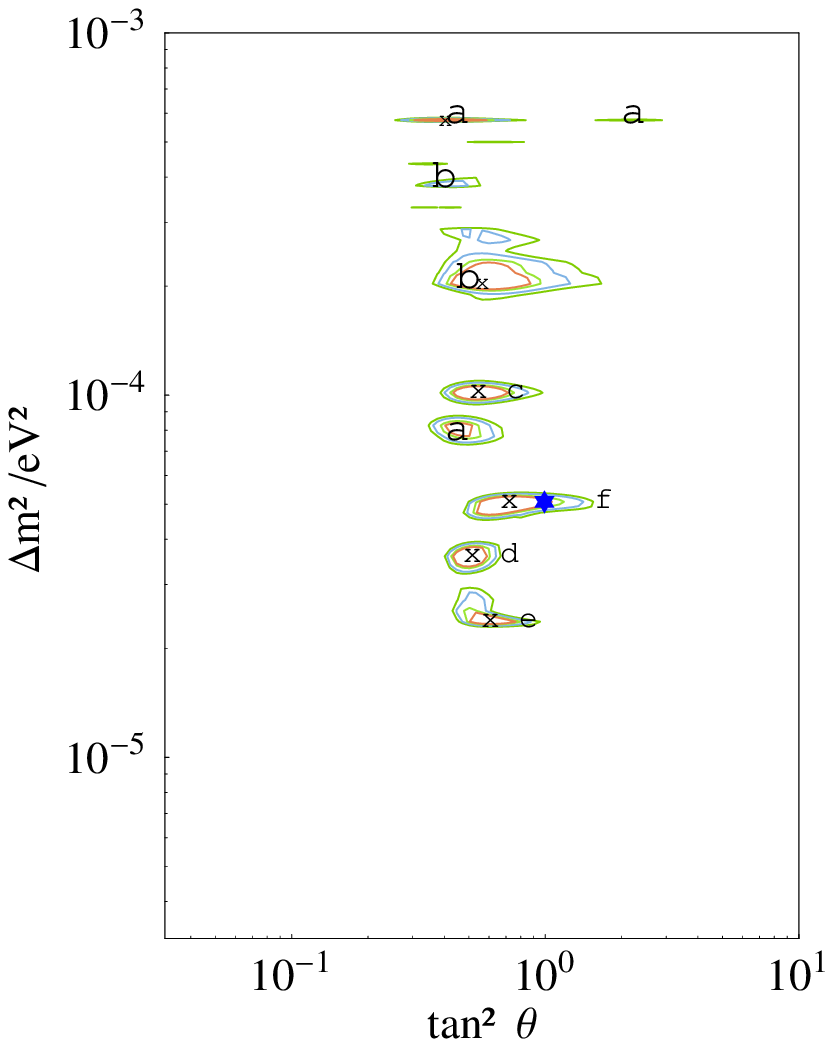,width=8cm}
\end{tabular}
\caption{\small 
Allowed areas in the two neutrino parameter space
after 3 years of data taking in KamLAND
(Analysis B).
Allowed regions belonging to the same point are labeled 
with the corresponding letter, the position of the 
point itself is labeled with a solid star
 (see table \protect\ref{t1}).
The colored lines separate allowed regions at 
90, 95, 99 and 99.7\% CL relative to the absolute minimum.
(Left) Results with the KamLAND spectrum alone.
(Right) KamLAND spectrum plus solar (CL,GA,SK,SNO) evidence.
Crosses are situated in the position of the $\chi^2$ minima.
}
\label{f3b}
\end{figure}

In Figs.(\ref{f3b})  we graphically 
show the results of this analysis for a selection of points
and for three years of data taking restricting ourselves 
to the LMA region of the parameter space where, as we have 
noted before, the KamLAND spectrum information is specially
significative.  
In each plot allowed regions corresponding to different 
starting points are superimposed, every region is 
distinguished with a label. The position of initial points is 
labeled with solid stars.

The first case, study of the KL spectrum alone
 ($\chi^2_{spec,KL}$) is represented by the 
 Fig.(\ref{f3b} left). 
The allowed parameter space corresponding to each 
particular point is formed by a number of,
 highly degenerated, disconnected 
regions symmetric with respect the line $\tan\theta=1$.
These regions can extend very far from the initial point 
specially in terms of $\Delta m^2$ but also in some 
occasions in terms of $\tan^2\theta$.
For example the point ``A'' located at 
$(\Delta m^2=5.7 \times 10^{-4},\tan^2\theta=0.38 )$
gives rise to two sets of thin regions situated  respectively
at $\Delta m^2\sim 10^{-3},10^{-4}$ and a third region 
situated at 
$\Delta m^2\sim 10^{-5}$ 
which practically covers the full range $\tan^2\theta\sim 0.1-10$.
A similar behavior is observed for point ``B''.
Of course this situation is not very  favorable for the 
future phenomenologist trying to extract conclusions 
from the KamLAND data. A much comfortable situation is 
found for points nearer the center of the LMA region.
Note how the regions corresponding to the points 
``D,E'' and specially ``F'' only extend very gently around 
the initial location.

The results of the full analysis are summarized in
 Fig.(\ref{f3b}, right). 
The position of the minima of $\chi^2$, marked in the 
plot with crosses, is practically identical to the position 
of the initial points except in some case where the difference
is not significant anyway.
The general effect of the inclusion of the solar 
evidence in the $\chi^2$ is the breaking of the symmetry in 
$\tan^2 \theta$ as expected and the general reduction 
 of the number of disconnected regions corresponding to 
each point. Note however that the point ``A'' still 
gives rise to a small allowed region situated nearly one
 order 
of magnitude smaller in $\Delta m^2$. The ``B'' region is 
shrinked near its initial location as happens  to the 
rest of points.
The conclusion to be drawed from these plots is that 
KamLAND together with the rest of solar experiments 
will be able to resolve the neutrino mixing parameters with
 a precision of $\delta \log \Delta m^2\sim \pm 0.1$ practically
 everywhere. However, for values of $\Delta m^2> 10^{-4}$
 the problem of the coexistence of multiple regions with 
 similar statistical significance will still be present.

\section{ Summary and  Conclusions}\label{sec:conclusions}
\label{conclusions}

We have analyzed the present experimental situation of our 
knowledge of the neutrino mixing parameters in the region of the 
parameter space that is relevant for solar neutrinos and 
we have studied in detail how this knowledge should improve 
with the forthcoming reactor experiment KamLAND.

In this work we have presented  in some detail the 
characteristic of 
KamLAND experiment including antineutrino reactor fluxes and 
absortion cross sections for which we have given updated 
values. We find that present theoretical errors in this 
cross secion are negligible.

We have  studied the expected KamLAND spectrum for different 
possible (``benchmark points'') values of the mixing parameters 
selected inside the LMA region.
The shape of the spectrum shows a significant dependence 
on the values of the mixing parameters.
The spectrum distortion caused by oscillation 
have been characterized by  the first moment of the 
positron energy distribution.
The results confirm that KamLAND is very sensitive 
to neutrino oscillation in the LMA region. 
In particular the value of the first moment changes very 
much for values of $\Delta m^2$ varying inside 
the region $\Delta m^2 \simeq 10^{-5} - 10^{-4} eV^2$. 
The dependence on the value of the mixing angle is also evident, 
even if somehow milder.
In the LOW and SMA  region, instead, 
the value of the moment is essentially constant.
We have also verified that the result doesn't depend in 
a significant way on the choice of the bin size.

In order to investigate the discrimination power of KamLAND, 
we have selected different points
(``benchmark points'')  in the LMA region and 
 studied which information KamLAND will be able to give 
after 1-3 years of running.
We have included a  
full modelling of statistical and systematic uncertainties.
The regions selected by KamLAND alone,  all symmetric with
 respect to $\tan^2 \theta =1$, have a large 
spreadth in the mixing angle.
The experiment should have, instead, a much higher sensitivity 
to the mass difference parameter, especially for values 
of $\Delta m^2$ lying in the central part of the LMA 
region. 
The situation would be more complicate 
for values closer to the border of the LMA region or beyond
already in the HLMA   region 
(i.e. $\Delta m^2 \leq 2 \times 10^{5}$
and $\Delta m^2 \geq 8-9 \times 10^{-5}$ or 
$\tan^2 \theta$ far from 0.5). 
In this case KamLAND, with or without solar evidence,
 will be able only to select 
multiple regions in the parameter space, in the sense that 
different possible 
values of the parameters would produce the same signal.

We have performed a similar analysis adding to 
the information from KamLAND would-be signal all the 
evidence already exististing on  solar neutrinos.
By using a $\chi^2$ analysis, we have produced exclusion 
plots.
KamLAND will help to select the values of the mixing
parameters especially in the case in which the solutions lies 
in the LMA region. If, instead, one moves towards values of 
the KamLAND signal closer
to the no oscillation value 
(that at present seems to be strongly disfavored by the other 
experiments) the absolute $\chi^2$ minimum moves from the LMA to 
the LOW solution and one is left with small allowed regions 
not only in the LOW, but also in the SMA region. 
In summary, KamLAND together with the rest of solar experiments 
will be able to resolve the neutrino mixing parameters with
 a high precision  practically
 everywhere. However, for values of $\Delta m^2> 10^{-4}$
(graphically emphatized 
by the two benchmark points labeled ``A'' and ``B'' in the plots)
 the problem of the coexistence of multiple regions with 
 similar statistical significance will still be present.

\vspace{0.3cm}
\subsection*{Acknowledgments}
It is a pleasure to thank R. Ferrari 
 for many enlightening discussions and 
 for his encouraging support.
We  acknowledge the  financial  support of 
 the Italian MURST and  the  Spanish CYCIT  funding 
agencies. One of us (E.T.) wish to acknowledge in 
addition the hospitality of the 
CERN Theoretical Division at the early stage of
this work. 
P.A., V.A., and M.P. would like to thank the 
kind hospitaliy of the Dept. de Fisica Teorica of the
U. Autonoma de Madrid.
The numerical calculations have 
been performed in the computer farm of 
 the Milano University theoretical group.

\newpage

\begin{table}
\centering
\begin{tabular}{lll}
\hline \hline
Label&$\tan^{2}\theta$&$\Delta m^{2}$ (eV$^{2}$)\\
\hline
A & $0.38$ & $5.70\times 10^{-4}$ \\
B & $0.60$ & $2.04\times 10^{-4}$ \\
C & $0.50$ & $1.01\times 10^{-4}$ \\
D & $0.50$ & $3.60\times 10^{-5}$ \\
E & $0.56$ & $2.37\times 10^{-5}$ \\
F & $0.99$ & $5.07\times 10^{-5}$ \\
\hline \hline\\
\end{tabular}
\caption{Benchmark points in the 2-neutrino parameter plane 
 used in the analysis.}
\label{t1}
\end{table}

\newpage

\end{document}